\documentclass[manuscript]{acmart}
\usepackage{amsmath,amsfonts}
\usepackage{algorithmic}
\usepackage{graphicx}
\usepackage{textcomp}
\usepackage{makecell}
\usepackage{subcaption}
\usepackage{listings}
\usepackage{fancyhdr}
\usepackage{multirow}
\usepackage{soul}
\definecolor{mygreen}{HTML}{2DA44E}
\usepackage{balance}
\usepackage{multicol}

\usepackage{tcolorbox} 

\newcommand{\rqboxc}[1]{\begin{tcolorbox}[left=3pt,right=3pt,top=3pt,bottom=3pt,colback=gray!5,colframe=gray!40!black,before skip=5pt,after skip=5pt]#1\end{tcolorbox}}

\newcommand{\phead}[1]{\vspace{1mm} \noindent {\bf #1}}
\newcommand{\uhead}[1]{\vspace{1mm} \noindent {\underline {#1}}}

\newcommand{\iuhead}[1]{\vspace{1mm} \noindent {\ul {\textit {#1}}}}

\newcommand{\inlineCode}{\lstinline[basicstyle=\normalsize\ttfamily]}

\usepackage{booktabs}

\definecolor{dkgreen}{rgb}{0,0.6,0}
\definecolor{gray}{rgb}{0.5,0.5,0.5}
\definecolor{mauve}{rgb}{0.58,0,0.82}
\definecolor{darkgreen}{rgb}{0.01, 0.75, 0.24}
\AtBeginDocument{%
  \providecommand\BibTeX{{%
    \normalfont B\kern-0.5em{\scshape i\kern-0.25em b}\kern-0.8em\TeX}}}

\setcopyright{acmcopyright}
\copyrightyear{2018}
\acmYear{2018}
\acmDOI{XXXXXXX.XXXXXXX}

\acmPrice{15.00}
\acmISBN{978-1-4503-XXXX-X/18/06}

\begin{document}

\title{Back to the Future! Studying Data Cleanness in Defects4J and its Impact on Fault Localization}

\author{Md Nakhla Rafi}
\email{r_mdnakh@encs.concordia.ca}
\affiliation{%
  \institution{Concordia University}
  \city{Montréal}
  \state{Québec}
  \country{Canada}
}

\author{An Ran Chen}
\email{anran6@ualberta.ca}
\affiliation{%
  \institution{University of Alberta}
  \city{Edmonton}
  \state{Alberta}
  \country{Canada}
}

\author{Tse-Hsun (Peter) Chen}
\email{peterc@encs.concordia.ca}
\affiliation{%
  \institution{Concordia University}
  \city{Montréal}
  \state{Québec}
  \country{Canada}
}

\author{Shaohua Wang}
\email{davidshwang@ieee.org}
\affiliation{%
  \institution{Central University of Finance and Economics}
  \city{Beijing}
  \country{China}
}

\renewcommand{\shortauthors}{Chen et al.}

\begin{abstract}
    For software testing research, Defects4J stands out as the primary benchmark dataset, offering a controlled environment to study real bugs from prominent open-source systems. However, prior research indicates that Defects4J might include tests added post-bug report, embedding developer knowledge and affecting fault localization efficacy. In this paper, we examine Defects4J's fault-triggering tests, emphasizing the implications of developer knowledge of SBFL techniques. We study the timelines of changes made to these tests concerning bug report creation. Then, we study the effectiveness of SBFL techniques without developer knowledge in the tests. We found that 1) 55\% of the fault-triggering tests were newly added to replicate the bug or to test for regression; 2) 22\% of the fault-triggering tests were modified after the bug reports were created, containing developer knowledge of the bug; 3) developers often modify the tests to include new assertions or change the test code to reflect the changes in the source code; and 4) the performance of SBFL techniques degrades significantly (up to --415\% for Mean First Rank) when evaluated on the bugs without developer knowledge. We provide a dataset of bugs without developer insights, aiding future SBFL evaluations in Defects4J and informing considerations for future bug benchmarks.
\end{abstract}





\maketitle
\section{Introduction}\label{sec:introduction}


Locating and fixing bugs is an expensive and time-consuming task, especially due to the ever-increasing complexity of modern software. A prior research~\cite{britton2013reversible} found that developers spend nearly 50\% of their working time on bug-fixing activities. Particularly, understanding the cause and locating the bug is often the most challenging step~\cite{parnin2011automated, wong2016survey, lin2017feedback, hailpern2002software}. Developers need to analyze all the available information, such as test failures and program execution details, to identify potential buggy code statements in the system and 
develop a bug fix. 


To assist developers in locating and fixing bugs, prior research has proposed techniques such as spectrum-based fault (SBFL) localization~\cite{pearson2017evaluating, sohn2017fluccs, zhang2017boosting, liu2019you, wang2016amalgam+} and automated program repair~\cite{icse20,li2022dear,saha2017elixir}. 
Given a set of failed tests, SBFL techniques analyze the test coverage to suggest a ranked list of possible buggy statements in the code. SBFL techniques rank the buggy statements based on the intuition that a statement covered by more failed tests and fewer passed tests is more likely to contain a bug. SBFL techniques are also an important building block for automated program repair techniques since they require potential buggy locations to start repairing~\cite{liu2019you}.


Given the increasing community attention on software testing research, prior studies introduced benchmarks for automatic program repair and fault localization. For example, there are many benchmarks available for both C and Java, such as ManyBugs~\cite{LeGoues15tse}
QuixBugs~\cite{lin2017quixbugs}, Bugs.jar~\cite{saha2018bugs}, Bugswarm~\cite{tomassi2019bugswarm}, and Defects4J~\cite{defects4j}. 
These benchmarks often provide test coverage, test failure information, and the corresponding bug fix. Hence, researchers can easily apply the proposed techniques for evaluation by using these benchmarks. 
Particularly, Defects4J is one of the most widely-used benchmarks~\cite{pearson2017evaluating, sohn2017fluccs, zhang2017boosting, martinez2017automatic, motwani2018automated}. Defects4J aims to provide a controlled environment to study bugs collected from real systems. Defects4J v2.0 contains 835 bugs from 17 open source Java systems. Each bug in the Defects4J benchmark is accompanied by at least one failing test that triggers the bug (i.e., {\it fault-triggering test}), where the tests and bug fixes are collected from the system development history. 

Although Defects4J has significantly helped advance software testing research, prior studies~\cite{liu2021critical, koyuncu2019ifixr, just2018comparing, kabadifuture} pointed out concerns that some tests in Defects4J may contain developer knowledge of the bugs.
As found by Liu et al.~\cite{liu2021critical}, some fault-triggering tests may be taken from a version of the system where the bug has already been resolved. These tests may provide hints to fault localization techniques about the location of the bug. Furthermore, a recent study by Kabadi et al.\cite{kabadifuture} emphasized the significance of differentiating between future test cases, which developers create after fixing bugs, and existing test cases, which were established prior. This distinction is critical since future test cases may over-aid automated program repair techniques due to their specificity. Therefore, having such developer knowledge in the tests may impact the evaluation of existing fault localization techniques. 
Another work by Just et al.~\cite{just2018comparing} compares the characteristics of developer-written tests to user-provided tests for fault localization and automated program repair in Defects4J. However, it is important to note that developer-written tests, which may or may not be fault-triggering tests, can be created either before, during, or even after the bug fix, leading to different implications due to the presence of developer knowledge.

In this paper, we conduct the first comprehensive study on the fault-triggering tests in Defects4J. We define \textit{fault-triggering tests} as the tests that fail when the bug is introduced~\cite{defects4j,wen2019exploring}. In other words, fault-triggering tests fail in the presence of a bug, and pass once the bug is fixed. The goal of this study is to provide insights into the prevalence of developer knowledge in fault-triggering tests and its impact on fault localization techniques. 
Firstly, we classify the tests in relation to developers' bug-fixing activities (e.g., a test is newly added or modified after a bug was reported), which may help future studies to better utilize Defects4J when evaluating SBFL techniques. We find that 77\% of the fault-triggering tests in Defects4J contain developer knowledge of the bug: 55\% of the tests were newly added after the bugs were fixed, and 22\% of the tests were modified after the bugs were reported for replication and regression testing purposes.  
Secondly, we analyze developers' modifications on the fault-triggering tests and identify different categories of reasons why developers modify them. We find that developers often add new assertions or modify the tests to cope with source code changes (i.e. 77\% of the analyzed cases), e.g., update the test after the bug was fixed. 
Finally, we study the impact of developer knowledge on the results of fault localization techniques. We apply state-of-the-art SBFL techniques on the commit before and after the bugs were fixed. 
Our experimental results show that the state-of-the-art spectrum-based fault localization techniques perform significantly worse in the absence of developer knowledge in the tests.


A major advantage of Defects4J is its controlled environment that allows researchers to reproduce the bugs easily. 
However, as we found, 55\% of the tests provided by Defects4J were not available when the bugs were first reported. 
Hence, 
it is yet unknown how fault localization techniques generalize to less-controlled settings. Practically, developers may use fault localization tools to gain insights into the debugging locations before they start to fix the bugs.
In addition to the above-mentioned findings, our work further contributes to Defects4J by providing an organized dataset that separates the bugs with and without the developer's knowledge in tests, and the coverage of the tests when the bugs were reported. This new dataset is valuable to 1) evaluate new and past fault localization and automated program repair techniques in a more practical setting (i.e., in the absence of developers' knowledge); 2) provide insights and guidance for future bug benchmarks; and 3) provide a better understanding on the developers' bug-fixing process to suggest more settings for evaluation.

The main contributions of this paper are:
\begin{itemize}
    \item We present a case study examining the prevalence of developer knowledge in Defects4J's fault-triggering tests. Our findings show that the majority of fault-triggering tests (77\%) contain developer knowledge of the bug.

    \item We manually studied and categorized the modification on fault-triggering tests. In total, we uncovered five categories, such as modifying tests to co-evolve with code and adding new assertions. We believe the uncovered categories may provide a starting point for understanding developers' bug-fixing process and assist in better utilizing Defects4J or creating/extending bug benchmarks. 

    \item We evaluate the impact of having developer knowledge in fault-triggering tests by comparing the results of SBFL techniques between the buggy version (before the bug was reported) and the version provided by Defects4J. We find that the fault localization results on the buggy version are significantly worse than that on the Defects4J version, with a degradation up to --415\% for Mean First Rank and --701\% for Mean Average Rank.


    \item We provide a discussion on the implications of our findings and highlight future research direction. We also made our dataset publicly available online~\cite{our_repo}, which includes a comprehensive list of the bugs that contain developer knowledge, the details of our manual study results, and the test result and coverage that we collected for the buggy commit (before the bug was reported).  

    


\end{itemize}

In summary, future research may use our annotated data to re-evaluate fault localization and automated program repair techniques. \textit{It is important to clarify that our intention is not to critique Defects4J negatively, but rather to objectively highlight and quantify its implications on fault localization.} Our findings may also help create more comprehensive bug benchmarks.

\phead{Paper Organization.} Section~\ref{sec:background} discusses the background and motivation of this study. Section~\ref{sec:empirical} presents our studied dataset and the motivation, approach, and results of the research questions. Section~\ref{sec:discussion} discusses the findings and highlights potential future research directions. Section~\ref{sec:related} presents related work to our study. Section~\ref{sec:threats} discusses threats to validity. Finally, Section~\ref{sec:conclusion} concludes the paper. 
\section{Background and Motivation}\label{sec:background}


The Defects4J v2.0.0 benchmark contains 835 bugs from 17 Java open source systems~\cite{defects4j}. All 835 bugs are extracted from different phases of software development, and the 17 projects span a wide range of domains and maturities. The objective of this benchmark is to facilitate research in software testing and debugging. Due to its ease of use and the realistic nature of the bugs, Defects4J has been widely used for conducting research in fault localization~\cite{pearson2017evaluating, sohn2017fluccs, zhang2017boosting, liu2019you, wang2016amalgam+}, automated program repair~\cite{yang2021were, jiang2019manual, martinez2017automatic, motwani2018automated}, and automated test generation~\cite{yu2019alleviating, gay2020defects4j, shamshiri2015automatically}.

In Defects4J, each bug comes with at least one failed test to ensure reproducibility. This failed test is known as the \textit{fault-triggering test}. As not all bugs are guaranteed to have a fault-triggering test at the time they are first uncovered, Defects4J utilizes an automated step to mine candidate fault-triggering tests from the bug fix and buggy commit of the system~\cite{defects4j}. Specifically, a fault-triggering test must deterministically pass on the fixed commit and fail on the buggy commit. Every bug and fault-triggering test is then manually examined to eliminate irrelevant code changes, such as the addition of new features. By default, Defects4J uses developer-written tests as fault-triggering tests to reproduce the bugs.

However, fault-triggering tests may be mined from the bug-fixing commit where developers may have already fixed the bug or were in the process of fixing it. This can include developer knowledge (i.e., added information about the bug that was not available at the time the bug was reported) in the tests. Table~\ref{fig:m_example} provides an excerpt of a bug, CLI-51, from the Defects4J benchmark with a test that has developer knowledge. CLI-51 is a bug from Commons-Cli where the code misinterpreted parameter values as new parameters. When developers first received this bug report, there was no test failure in the system. Developers provided a patch to fix the bug and commented in the report that a fault-triggering test (i.e., {\sf BugCLI51Test}) was developed as part of the patch for regression testing. The fault-triggering test was introduced after the corresponding bug fixes. In particular, developers developed the test based on the content of the bug report and the bug fix. The fault-triggering test verifies the parameter value ``{\sf -t -something}'' that triggers the bug, as mentioned in the bug report.
As a result, the test contains developer knowledge, which provides hints on the causes and location of the bug in the source code.

\begin{table}[t]
\centering
\caption{An excerpt of the Bug report CLI-51 from the Defects4J benchmark.}
\resizebox{\columnwidth}{!}{
\begin{tabular}{p{.2\columnwidth}p{.9\columnwidth}}
    \toprule
    \textbf{BugID} & CLI-51\\
    \midrule
    Summary & Parameter value ``-something'' misinterpreted as a parameter \\
    \midrule
    Developer's comment & ``{\em Fix so parser doesn't burst options which are not defined. (-s) in the above case.} \\
    &{\em Includes unit test [BugCLI51Test].}''\\
    \bottomrule
\end{tabular}
\label{fig:m_example}
}
\vspace{-2em}
\end{table}

Although prior studies~\cite{koyuncu2019ifixr, wen2019historical, chen2022useful} suggest that some tests in Defects4J may contain developer knowledge, there is no systematic study on the prevalence of such tests and their impact on the results of fault localization techniques. Fault localization is an important step in assisting developers locate faults~\cite{pearson2017evaluating, zou2019empirical, lou2021boosting, b2016learning} and in guiding automated program repair techniques~\cite{le2016history, liu2019you, nguyen2013semfix}. Under- or over-estimating the effectiveness of fault localization techniques may affect their adoption in practice. Hence, in this paper, we study the fault-triggering tests in Defects4J in relation to the bug fixes, and their impact on the fault localization results. 
Our experimental results reveal that a majority of the fault-triggering tests (77\%) are affected by developer knowledge. We provided a classification of the bugs in Defects4J based on our findings and future research may consider our dataset when evaluating fault localization techniques. 




\section{Study Design and Results}\label{sec:empirical}
In this section, we first discuss an overview of the studied systems. Then, we present the motivation, approach, and results of the three research questions (RQs) that we seek to answer. 

\subsection{Overview of the Studied Systems}

\noindent We performed our study on the Defects4J (V2.0.0) benchmark~\cite{defects4j}, which includes real and reproducible faults from a wide range of systems. Defects4J forms the basis of many prior studies on fault localization~\cite{wen2019historical, pearson2017evaluating, zhang2017boosting, li2019deepfl, sohn2017fluccs, liu2019you}, where these studies use Defects4J as the benchmark dataset for evaluation and comparison with state-of-the-art. 
Table~\ref{tab:overview} provides an overview of our studied systems. We collected the lines of code, LOC, and number of tests from the HEAD version of each system. The sizes of the studied systems in Defects4J range from 4K to 90K lines of code, and the benchmark contains 1,655 fault-triggering tests.
We did not consider the system Chart from Defects4J since it does not use Git as the version control system, and we rely on analyzing the development history in Git repositories to understand the changes on the fault-triggering tests for our study.
In total, we conducted our experiments on 809 bugs from 16 systems. Note that, one bug may have more than one fault-triggering test, so there are more fault-triggering tests than the number of bugs. 


\begin{table}[t]
\caption{An overview of our studied systems from Defects4J}
\centering
\resizebox{.6\columnwidth}{!}{
\begin{tabular}{llrrr}
    \toprule
    \textbf{System} & \textbf{\#Bugs} & \textbf{LOC} & \textbf{\#Tests} & \textbf{Fault-triggering}\\
    &&&&\textbf{Tests}\\
    \midrule
    Cli          & 39       & 4K     & 94     &  66      \\
    Closure      & 174      & 90K    & 7,911  &  545     \\
    Codec        & 18       & 7K     & 206    &  43      \\
    Collections  & 4        & 65K    & 1,286  &  4       \\
    Compress     & 47       & 9K     & 73     &  72      \\
    Csv          & 16       &  2K    & 54     &  24      \\
    Gson         & 18       & 14K    & 720    &  34      \\
    JacksonCore  & 26       & 22K    & 206    &  53      \\
    JacksonDatabind & 112   & 4K  & 1,098  &  132     \\
    JacksonXml   & 6        & 9K     & 138    &  12      \\
    Jsoup        & 93       & 8K     & 139    &  144     \\
    JxPath       & 22       & 25K    & 308    &  37      \\
    Lang         & 64       & 22K    & 2,291  &  121     \\
    Math         & 106      & 85K    & 4,378  &  176     \\
    Mockito      & 38       & 11K    & 1,379  &  118     \\
    Time         & 26       & 28K    & 4,041  &  74      \\
    \midrule
    \textbf{Total}& 809     & 409K & 25,708 & 1,655 \\
    \bottomrule
\end{tabular}
\label{tab:overview}
}
\end{table}

\subsection{RQ1: Were Fault-triggering Tests Added/Modified After a Bug Was Reported?}

\phead{Motivation.} 
When localizing faults, SBFL techniques rely mainly on analyzing the coverage of the fault-triggering tests. 
However, prior research~\cite{koyuncu2019ifixr, chen2022useful, wen2019historical} suggests that Defects4J may contain some tests that were added by developers 
{\em after the bug was reported or fixed}.   
Such tests may contain developer knowledge of the bug, which compromises the reliability of the results of fault localization techniques.
Therefore, in this RQ, we study the timelines of 
test modification/addition for every bug, starting from the creation of the bug report until its resolution. 
We also study whether or not the changes in tests bring developer knowledge of the bugs to the test. 
The findings will provide initial evidence on how many bugs from Defects4J whose fault-triggering tests may have developer knowledge. 

\begin{table*}
\caption{Timelines of the changes on the fault-triggering tests. Note that the same bug may belong to more than one pattern because a bug may have more than one fault-triggering test.}
\centering
\resizebox{.8\textwidth}{!}{
    \begin{tabular}{p{1.2cm}p{5cm}>{\raggedleft\arraybackslash}p{6cm}cc}

        \toprule
        \textbf{Patterns} & {\bf Description} & \multicolumn{1}{c}{\textbf{Example Timeline}} & \textbf{\#Tests} & \textbf{\#Bugs}\\
        \midrule
        Pattern 1 & Fault-triggering tests were newly added after the bug was reported. &
        \begin{minipage}{.37\textwidth}
          \includegraphics[width=1.05\linewidth, height=26mm]{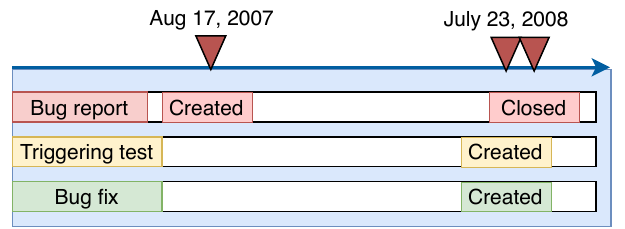}
        \end{minipage}
        & 872 (53\%) & 558\\\\ \hline
        Pattern 2 & Fault-triggering tests were newly added then modified, after the bug was reported. & 
        \begin{minipage}{.37\textwidth}
          \includegraphics[width=1.05\linewidth, height=26mm]{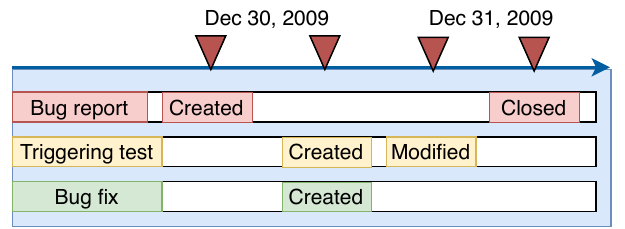}
        \end{minipage}
        & 43 (2\%)  & 30    \\\\ \hline
        Pattern 3 & Fault-triggering tests were modified after the bug was reported and before the bug was marked as fixed. &
        \begin{minipage}{.37\textwidth}
          \includegraphics[width=1.05\linewidth, height=26mm]{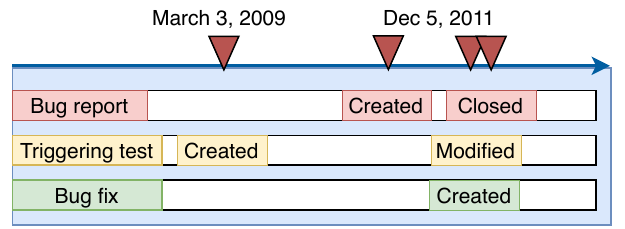}
        \end{minipage}
        & 362 (22\%) & 155   \\\\ \hline
        Pattern 4 & Fault-triggering tests were not modified after the bug was reported and before the bug was marked as fixed. & 
        \begin{minipage}{.37\textwidth}
          \includegraphics[width=1.05\linewidth, height=26mm]{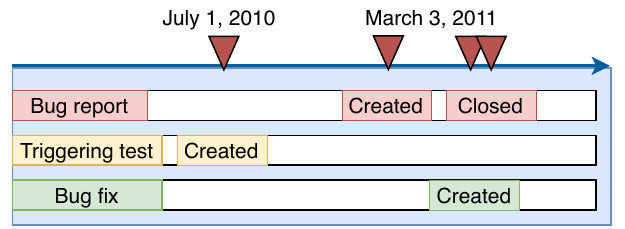}
        \end{minipage}
        & 378 (23\%) & 150   \\\\
        \midrule
        \textbf{Total} &  & &1,655 & -\\
        \bottomrule
    \end{tabular}
    \label{tab:patterns}
}
\vspace{-1em}
\end{table*}

\phead{Approach.} We conducted a tool-assisted manual study on the timelines of events for every bug. We first collected the bug report, fault-triggering tests, and bug-fixing patches for all of the 809 studied bugs. We then identified the events (e.g., bug report creation) associated with each piece of information. We analyze these events automatically in relation to bug resolution and manually reviewed the test modifications to ensure their relevance to the bug. Below, we discuss our data collection process in detail.

\iuhead{Bug report creation and resolution date:} To determine the time interval of the bug resolution and identify whether the fault-triggering tests were modified during this period, we collected the creation and resolution time of the bug reports. We retrieved the bug reports from the bug tracking systems (i.e., Jira and GitHub) using REST APIs and the bug IDs provided in Defects4J. We then extracted the creation and resolution time from the ``Created'' and ``Resolved'' fields (or the issue creation and closed dates on GitHub) of each bug report. By using the creation and resolution date, we are able to determine if a test was added/modified after a bug was reported or fixed. 


\iuhead{Date of fault-triggering test creation and modification:} 
We identified all the commits that are associated with the fault-triggering tests and analyzed when the commits happened (e.g., before or after the bug was reported/fixed). 
We used the git command ``\inlineCode{git log -L:[funcname]:[file]}'' to identify the list of commits that modified the fault-triggering test and the modification date. 

\iuhead{Bug fix date:} 
In addition to the bug report creation/resolution time, we study the time of the bug fix to understand if a test was modified before or after the fix became available. 
We used the ``\inlineCode{git log [commit]}'' command to determine the exact date and time of the bug fixes. We then align the time of the bug fixes with modifications made to the fault-triggering tests. 

\iuhead{Change patterns of the tests and their relevance to the bugs:}
We arranged the events -- bug report creation and resolution, creation and modification of fault-triggering tests, and bug fix time -- in chronological order to reconstruct the timeline. In particular, we focused on the creation and modifications of fault-triggering tests with respect to bug resolution.
We used an automated script to sequence the events into timelines. 
We then manually studied
the commit messages and related code changes to examine if the added/modified tests include developer knowledge about the bug. 
Our collected data is publicly available~\cite{our_repo}.

\phead{Results.} \textbf{\textit{We find that 77.2\% of the fault-triggering tests in Defects4J include developer knowledge of the bugs, bringing hints on the groundtruth of the buggy location in the data. }}
In total, we uncovered four timelines of change patterns on the fault-triggering tests. Table~\ref{tab:patterns} shows the uncovered change patterns and corresponding timelines of the events. In summary, through our manual analysis, we find that developers' modification or addition of tests adds bug-related information to all of the tests from Pattern 1 and Pattern 2, and most of the tests from Pattern 3 (357/362). Below, we discuss each pattern in detail.

\iuhead{Pattern 1: Test created after bug report creation (53\%).} When developers are trying to fix a bug, they often rely on fault-triggering tests to understand and replicate the problem~\cite{cotroneo2013fault, wen2019exploring, sohn2021leveraging, chen2022useful, an2022fonte}. However, as illustrated in Table~\ref{tab:patterns}, we found that in Defects4J, a large portion of these fault-triggering tests may not exist before the bug report was created. The most common change pattern is that the tests were added to the system only after developers began to fix the bug. 
As an example, in CLI-144, the bug report was created on August 17, 2007. On July 23, 2008, developers created a new fault-triggering test called {\sf BugCLI144Test} as part of the bug fixing commit (the test was named using the bug report ID). This new fault-triggering test was added because existing tests were unable to capture the reported bug. Thus, the test contains information on the groundtruth of the bug-fixing location. However, this test was marked as the fault-triggering test in Defects4J. 
Through manual inspection, we found that all the fault-triggering tests that belong to this pattern were added by developers as part of the bug-fixing process. Namely, nearly 50\% of the fault-triggering tests in Defects4J were regression tests that were added deliberately to replicate/prevent the bug. 


\iuhead{Pattern 2: Test created and modified after bug report creation (2\%).} As shown in Table~\ref{tab:patterns}, after a bug report is created, developers may create initial versions of the fault-triggering tests that require further enhancement (e.g., the initial version is not able to cover all possible scenarios). Nevertheless, in our manual analysis, similar to Pattern 1, we found that these fault-triggering tests were created for regression testing purposes and contained developer knowledge of the bug.  
As an example, in MATH-320, the developer initially added a bug fix and a new test to reproduce the bug. However, the developer commented that the fix was incomplete and shared the test to facilitate discussions with other developers. Later, the developer applied a patch to fix the bug along with the updated test. 
For all the tests that belong to Pattern 2, either the commit messages or the names of these fault-triggering tests contain the ID of the bug report, further confirming that they contain developer knowledge of the bug.    


\iuhead{Pattern 3: Test modified after bug report creation (22\%).} In practice, some bugs reported by users or other developers cannot be revealed by existing tests. Hence, when fixing these bugs, developers may modify and enhance the tests for regression testing purposes. For example, in COMPRESS-10, the test {\sf UTF8ZipFilesTest} was enhanced as part of the bug fix. Developers modified the assertions to better capture the bug. 
It is possible that the test modification is not related to the bug fix. However, after our manual analysis, we found that 99\% (357/362) of the tests contained changes that alter the test execution for replicating the bug, while the remaining 1\% (5/362) did not introduce new knowledge to the tests (e.g., code re-styling, and enabling or disabling a failed test). In short, we find that most modifications to the fault-triggering tests were done after the bug was reported and were adding developers' knowledge of the bug in the tests. 



\iuhead{Pattern 4: Test unmodified during bug resolution (23\%).} We found that developers may fix the bug without making any changes to the fault-triggering tests. As an example (CLOSURE-79), when the problem was initially uncovered, the fault-triggering test {\sf testPropReferenceInExterns3} failed. Developers work on the bug fix without having to change the test as it was working as intended (i.e., failing upon unexpected behaviour). The fault-triggering tests were not changed during the resolution of the bug report. In total, we found 378 manually-verified fault-triggering tests that belong to this pattern.

\phead{Discussion.}
In our manual analysis, we identified four common change patterns on the fault-triggering tests. Based on the results of Pattern 1 and 2, {\it we observed that 55\% (915/1,655) of the fault-triggering tests were created after the bug report was created}. 
These tests did not exist before the bugs were reported, so they could not have helped in identifying the bugs. Moreover, these tests contain developers' knowledge of the bug, adding hints on the groundtruth of the bug location. 
Leveraging these tests in fault localization introduces biases in how well code coverage can help locate the faults.
We also observed that developers may modify the tests after the creation of the bug report, which is the case in 22\% (362/1,655) of the studied fault-triggering tests. This can lead to a similar problem where tests were modified with bug specifications (i.e., developer knowledge). In particular, the modified version of the test can differ significantly from the initial version, and it may be modified to specifically address the problem described in the bug report. 
In short, we find that only 23\% of the studied fault-triggering tests were actually able to detect the bugs (i.e., the tests failed), which account for 19\% (150/809) of the total bugs. 

\rqboxc{{\bf RQ1-Takeaway.} The majority of the fault-triggering tests in Defects4J (77\%) contain developer knowledge. These tests were either newly added or modified to replicate the bug or to prevent future regression. Only 23\% of the tests were able to detect the bugs (19\% of the total bugs) as intended.}

\subsection{RQ2: How Do Developers Modify the Fault-triggering Tests?}
\begin{table*}[t]
\centering
\caption{List of categories of the modifications to fault-triggering tests.}
\resizebox{.9\textwidth}{!}{
    \begin{tabular}{p{3cm}p{12cm}rr}
    \toprule
    \textbf{Category} & \textbf{Description} & \textbf{Count} & \textbf{Percentage} \\ \midrule

     Adding New Test & Developers added a new test to reproduce the bug. & 189 & 63\% \\
    Test Co-evolution & Developers modified the expected output in tests to cope with source code evolution. & 58 & 20\% \\
    Adding New Assertion & Developers added a new assertion to replicate the bug and for regression testing purposes. & 41 & 13\% \\
    Improving Test Logic During Bug Fixes & While modifying a test to replicate the bug, developers also enhance the structure or the logic in the test.  & 7 & 2\% \\
    Others  & Developers reformated the style in the test (e.g., indentation), or eliminated code that is not essential to the reproduction of faults. & 2 & 1\%\\
    
    \bottomrule
    \end{tabular}
    \label{tab:category}
}
\vspace{-1em}
\end{table*}

\phead{Motivation.}
When fixing a bug, developers may look for test failures to help in debugging and understanding the root causes of the bug~\cite{jiang2009well, kim2013should, labuschagne2017measuring, chen2022t}. However, as we found in RQ1, many fault-triggering tests may not exist before the bug report. Even when they do exist, they can only trigger the bug after developers made modifications to them during the bug-fixing process.
In this RQ, we manually study the modifications made by developers to the fault-triggering tests and discuss the common reasons behind them. Studying why developers modify these tests can help us understand what motivates them to make changes and how these changes relate to fixing the bugs.

\phead{Manual Study Process.}
We conduct our manual study on all 16 studied systems. For a 95\% confidence level and a 5\% confidence interval, we randomly sample 300 fault-triggering tests from 1,361 modified tests that were collected from Patterns 1, 2, and 3.
We use the stratified sampling strategy~\cite{neyman1992two} to obtain the number of samples for each studied system that is proportional to their total number of tests.
For each test, we study its code changes, code comments, commit message, and corresponding bug report and bug fix to understand the potential motivation leading to its modification. Following prior studies~\cite{li2019dlfinder, lamothe2020apis, ding2020towards, li2020shall}, we conduct our manual study in three phases using the open coding method.

\iuhead{Phase 1:} The first author manually reviews the tests to create a preliminary list of categories for 100 fault-triggering tests that are randomly selected. The author additionally lists the justifications for creating each category in terms of its code changes, code comments, commit message, bug report, and bug fix. Together with the second author, the two authors revise the list of categories and address any discrepancies.

\iuhead{Phase 2:} The two authors review the remaining of the 300 tests independently based on the discussed categories and assign the test to one of the uncovered categories from Phase 1. 

\iuhead{Phase 3:} The two authors compare their assigned categories and discuss any disagreements until they reach a consensus. During this phase, our results show that the authors reached a substantial level of agreement, achieving a Cohen's Kappa of 0.86~\cite{cohen1960coefficient}.

\phead{Results.} \textbf{\textit{Other than test addition, most changes on the fault-triggering tests are related to improving test oracles or updating tests in response to source code changes.}}
In our manual analysis, we uncover five categories of reasons why developers changed the fault-triggering tests as shown in Table~\ref{tab:category}. 
Below, we discuss each category in detail. 


\iuhead{Adding New Test (189/300, 63\%).} We found that some tests are specifically created to aid in the reproduction of a bug. As developers work towards resolving a bug, they may create new tests designed to replicate the problematic behavior. Once the bug is fixed, developers often incorporate these tests into their patch to ensure that the bug does not re-occur in future releases (i.e., regression testing). All test changes from this category belong to Patterns 1 and 2 that we found in RQ1. These tests are often named after the bug report ID to ease traceability and management. 

\iuhead{Test Co-evolution (58/300, 19\%).}
We find that developers may change the test code to accommodate the bug-fixing changes in the source code. 
As shown in Listing~\ref{lst:test_coevolution}, developers fixed bug \#207 in JacksonCore by modifying the {\sf calcHash} method in the source code (as shown below). In addition to the bug fix, developers patched the test {\sf testSyntheticWithBytesNew} responsible for verifying the stability of the hash code, which failed after the new bug fix. One of the developers highlighted in a comment that this bug fix improved ``{\em hashing [with regards to] existing test}''. The system is expected to have a different distribution of collision count. Therefore, developers modified the test to match its expected output to the actual behavior of the system.

\begin{lstlisting}[caption={Example of test co-evolution.},captionpos=b,label={lst:test_coevolution},language=Java]
    // ByteQuadsCanonicalizer.java
    public int calcHash(int q1){
        int hash = q1 ^ _seed;
        hash += (hash >>> 16); // to xor hi- and low- 16-bits
        (*@\color{red}- hash \^{}= (hash >>> 12);@*) // as well as lowest 2 bytes
        (*@\color{green}+ hash \^{}= (hash << 3);@*) // shuffle back a bit
        (*@\color{green}+ hash += (hash >>> 12);@*) // and bit more
        return hash;
    }
    
    // TestSymbolTables.java
    public void testSyntheticWithBytesNew() throws IOException{
        ...
        // anywhere between 70-80% primary matches
        (*@\color{red}- assertEquals(8524, symbols.primaryCount());@*)
        (*@\color{green}+ assertEquals(8534, symbols.primaryCount());@*)
    }
\end{lstlisting}
\vspace{-1.5em}
    


\iuhead{Adding New Assertion (41/300, 14\%).} During the bug-fixing process, developers may add new assertions to help reproduce a bug (belongs to Pattern 3 from RQ1). Typically, as we found in our manual study, the modification to the test involves adding single-line assertion statements. For example, as seen in the test {\sf testCreateNumber} shown in Listing~\ref{lst:new_assert} (LANG-822), developers added a new assertion to reproduce the bug. The assertion checks whether the method {\sf createNumber} can execute as expected if the input starts with the string ``\lstinline{--}''. 
According to the developer's comment, there are numerous edge cases that can expose problematic behaviors when calling the method, so whenever a new edge case arises, it is added to the test as a new assertion statement.

\begin{lstlisting}[caption={Examples of new assertion.},captionpos=b,label={lst:new_assert},language=Java]
    public void testCreateNumber() {
    ...
       // LANG-693
       assertEquals("createNumber(String) LANG-693 failed", Double.valueOf(Double.MAX_VALUE), NumberUtils.createNumber("" + Double.MAX_VALUE));
    (*@\color{darkgreen}+ // LANG-822@*)
    (*@\color{darkgreen}+ assertFalse("createNumber(String) LANG-822 succeeded", checkCreateNumber("--1.1E-700F"));@*)
    }
\end{lstlisting}
\vspace{-1em}

\iuhead{Improving Test Logic During Bug Fixes (7/300, 2\%).}
Developers may modify tests to change the logic of the tests when trying to fix a bug. Typically, these modifications happen when developers are trying to replicate the bug in a test. Examples of modifications include simplifying complex logic and reorganizing the code structure, which can alter the execution of the test. In Figure~\ref{lst:testlogic}, we show an example of a test modification in this category based on the {\sf JsonAdapterNullSafeTest} from issue \#800 in GSON. During the bug-fixing process, developers discussed providing a ``simpler'' test to reproduce the bug. They incorporated new changes that simplified the initialization of the {\sf Device} class from the test and removed logic that was irrelevant in triggering the fault. This contributed to improving the quality and maintenance of the test.

\begin{lstlisting}[caption={Example of improved test logic during bug fixes.},captionpos=b,label={lst:testlogic},language=Java]
    public void testNullSafeBugDeserialize() throws Exception {
      (*@\color{red}- String json = "\textbackslash"{\textbackslash\textbackslash\textbackslash"id\textbackslash\textbackslash\textbackslash":\textbackslash\textbackslash\textbackslash"ec57803e2\textbackslash\textbackslash\textbackslash",\textbackslash\textbackslash\textbackslash"category\textbackslash\textbackslash\textbackslash":2}\textbackslash"";@*)
      (*@\color{red}- Device device = gson.fromJson(json, Device.class);@*)
      (*@\color{darkgreen}+ Device device = gson.fromJson("{'id':'ec57803e2'}", Device.class);@*)
      ...
      @JsonAdapter(Device.JsonAdapterFactory.class)
      private static final class Device {
        String id;
        (*@\color{red}- int category;@*)
        (*@\color{red}- Device(String id, int category) {@*)
        (*@\color{darkgreen}+ Device(String id) {@*)
          ...
      }
\end{lstlisting}
\vspace{-1.5em}

\iuhead{Others (2/300, 1\%).}
We find two other reasons why developers may modify the tests, which do not belong to any of the above categories. In particular, we observe that developers may reformat the source code files without necessarily changing any of the logic in the code. For instance, developers removed the extra indentation in the test to improve its readability. We also observe that developers may clean up the test which entails eliminating any obsolete variables or comments.

\rqboxc{{\bf RQ2-Takeaway.} Developers often add new tests (63\%) or assertions (14\%) to replicate the bug, and sometimes improve tests (21\%), e.g., reflecting the fix in source code or improving test logic, to regression test the bugs that they are fixing. }

\subsection{RQ3: How does Having Developer Knowledge in Fault-triggering Tests Affect Fault Localization Results?}

\phead{Motivation.} In the previous RQs, we found that many tests are newly added or modified for replicating the bugs and regressing testing. Such tests contain developer knowledge of the bugs that may affect the effectiveness of fault localization techniques. Therefore, in this RQ, we study the impact of using developer-modified tests for fault localization. The findings will provide insight into the impact of using these tests for fault localization, and whether there is a need to improve the benchmark. 



\phead{Approach.} 
Among existing fault localization techniques, spectrum-based fault localization (SBFL) is one of the most widely studied techniques as it provides an accurate ranking of the potentially buggy statements~\cite{li2019deepfl, abreu2006evaluation, keller2017critical}. SBFL techniques are also an important building block for automated program repair techniques since they rely on SBFL to provide a list of potentially buggy locations to start repairing~\cite{liu2019you}. SBFL techniques differentiate the buggy statements from the non-buggy ones through the program spectrum (code coverage profile). Intuitively, a statement covered by more failed tests but fewer passed tests is more likely to contain the bug. Therefore, developer-modified fault-triggering tests can significantly impact the performance of SBFL, as these tests may not have existed when the bug was first reported, or might not have been able to trigger the bug (i.e., do not fail) when initially introduced to the system.

In particular, we study the effectiveness of SBFL techniques on the bugs whose fault-triggering tests existed before the bug report and were modified after the bugs were reported. This is done to enable a comparison of fault localization techniques with and without developer knowledge of the bug.
The tests that belong to Pattern 3 existed before the bug report and were modified after the bugs were reported, and we found that most of the modifications added developer knowledge of the bug to the tests (RQ2). Therefore, we perform our study on bugs that belong to Pattern 3. 
To study the impact of including developer knowledge on SBFL, we compare the performance of SBFL techniques in two versions: 1) {\em vBuggy:} when the bug was first reported, and 2) {\em vD4J:} the version provided by Defects4J (the tests were modified during bug fixes). We perform our comparison on all 155 bugs that contain fault-triggering tests belonging to Pattern 3. 

Since Defects4J only provides the test coverage and test execution result for {\em vD4J}, we need to collect the data for {\em vBuggy}. To collect {\em vBuggy}, for each bug, we extract the timestamp when the bug report was created and identify the nearest commit {\em before} the bug was reported. In practice, the bugs remain unfixed in these commits, and if there are any test failures, they are more likely to be related to the bug. We checkout these commits on Git, compile the systems, and execute the tests. Note that, if a bug does not have any failed tests in {\em vBuggy}, we exclude the bug from our analysis for a more direct comparison between the two versions.
Since most systems do not report test coverage, we manually modified the systems to use JaCoCo to collect the coverage information. In total, we spent hundreds of hours configuring JaCoCo and resolving compilation issues such as missing dependencies and incompatible environment settings. We release the data publicly to encourage replication and future research~\cite{our_repo}. 

After collecting the data for {\em vBuggy}, we apply SBFL techniques on both {\em vBuggy} and {\em vD4J} and compare the localization result. In particular, we use four following commonly used SBFL techniques~\cite{GZoltar}: Ochiai~\cite{abreu2006evaluation}, Tarantula~\cite{Tarantula}, DStar~\cite{dstar}, and Barinel~\cite{Barinel}. To evaluate the fault localization techniques, we use the following metrics:

\uhead{{\em Top-K}} is defined as the number of faults with at least one faulty statement correctly identified within the first K statements in the ranking. Therefore, a better Top-K result indicates that developers are able to find faulty statements by examining fewer statements. We set K to 1, 3, and 5 in our evaluation. 

\uhead{{\em Mean First Rank (MFR)} } calculates, for all the bugs, the mean of the first faulty statement in the ranked result. Therefore, a smaller value means that the technique, on average, is able to identify a faulty statement early in the ranked list. 

\uhead{{\em Mean Average Rank (MAR)}} first calculates the average rank of all the faulty statements for a bug. Then, MAR calculates the average of the ranks from all the bugs. A smaller value means that the faulty statements are ranked earlier. 

\begin{table}[t]
\caption{Fault localization results for the bugs that belong to Pattern 3 (fault-triggering tests exist but were modified). We compare the results of running the techniques in the commits before the bugs were reported ({\em vBuggy}) and in the commits that were provided by Defects4J ({\em vD4J}). 
}
\label{tab:localizationPattern3}
\resizebox{.6\columnwidth}{!}{
\begin{tabular}{l|l|r|r|r|r|r|r}
\toprule
\textbf{Technique}         & \textbf{Version}  & \textbf{Bugs} & \textbf{Top-1} & \textbf{Top-3} & \textbf{Top-5} & \textbf{MFR} & \textbf{MAR} \\ \midrule
\multirow{2}{*}{Ochiai}    & {\em vBuggy} & 118            & 3              & 3              & 3              & 2,965 (-415\%)      & 3,254  (-268\%)     \\ 
                           & {\em vD4J}  & 118            & 49             & 64             & 70             & 714       & 1,214       \\ \midrule
\multirow{2}{*}{Tarantula} & {\em vBuggy} & 118            & 4              & 4              & 4              & 2,988  (-424\%)    & 3,244   (-262\%)     \\ 
                           & {\em vD4J}  & 118            & 47             & 62             & 66             & 705        & 1,238     \\ \midrule
\multirow{2}{*}{D-Star}   & {\em vBuggy} & 118            & 10              & 11              & 15              & 2,827 (-104\%)      & 3,231   (-701\%)   \\ 
                           & {\em vD4J}  & 118            & 29             & 39             & 44             & 273      & 461        \\ \midrule
\multirow{2}{*}{Barinel}   & {\em vBuggy} & 118            & 3              & 3              & 3              & 2,954 (-378\%)     & 3,282  (-272\%)     \\ 
                           & {\em vD4J}  & 118            & 45             & 60             & 65             & 781       & 1,208      \\ \bottomrule
\end{tabular}
}
\end{table}

\phead{Results.} \textbf{\textit{For all the four fault localization techniques that we studied, the localization results degrade significantly on vBuggy compared to vD4J on bugs in Pattern 3.}} Table~\ref{tab:localizationPattern3} shows the fault localization (FL) results on {\em vBuggy} and {\em vD4J} for the bugs whose fault-triggering tests belong to Pattern 3 (fault-triggering tests exist but were modified after the bug was reported). 
Out of the 155 unique bugs that we considered, 118 of them caused test failures. 
The remaining 25 bugs either did not lead to any test failures or their commits before the bug report ({\em vBuggy}) were too old and could not be retrieved. Thus, we perform our analysis on 118 bugs.
Table~\ref{tab:localizationPattern3} shows that the results in {\em vBuggy} are significantly worse than that of {\em vD4J} for all the metrics. All four FL techniques in {\em vBuggy} have much fewer times of a faulty element ranked in Top-1, Top-3, and Top-5 than that in {\em vD4J}. 
The finding indicates that in the best scenario, 15 of the bugs have faulty statements that are ranked early in the list. 
In comparison, for {\em vD4J}, the number of faulty elements being ranked in Top-1 and Top 5 is around 40 and 60, respectively. 
The MFR and MAR results for {\em vBuggy} are in the range of 3,000, whereas the results for {\em vD4J} are in the range of 200 to 1200. In other words, most faulty statements are ranked very low in {\em vBuggy} and the ranking results cannot help developers with locating the bugs. The decrease in MRF and MAR value are in the range of 100\% to 700\%.
In short, the findings indicate that, in {\em vBuggy}, the fault-triggering tests (without Pattern 3 tests having prior developer knowledge) are not able to help locate the bugs.

\begin{table}
\caption{Fault localization results for the bugs that belong to Pattern 4 (the fault-triggering tests were not modified). We compare the results of running the techniques in the commits before the bugs were reported ({\em vBuggy}) and in the commits that were provided by Defects4J ({\em vD4J}).} 
\label{tab:flPattern4}
\resizebox{.6\columnwidth}{!}{
\centering
\begin{tabular}{l|l|r|r|r|r|r|r}
\toprule
\textbf{Technique}         & \textbf{Version}  & \textbf{Bugs} & \textbf{Top-1} & \textbf{Top-3} & \textbf{Top-5} & \textbf{MFR} & \textbf{MAR} \\ \midrule
\multirow{2}{*}{Ochiai}    & {\em vBuggy} & 105           & 2              & 2              & 6              & 2,097  (-245\%)     & 2,561   (-203\%)       \\ 
                           & {\em vD4J}  & 105           & 10             & 18             & 26             & 855        & 1,264    \\ \midrule
\multirow{2}{*}{Tarantula} & {\em vBuggy} & 105           & 2              & 2              & 6              & 2,100 (-247\%)       & 2,587  (-206\%)       \\ 
                           & {\em vD4J}  & 105           & 12             & 17             & 28             & 851       & 1,254    \\ \midrule
\multirow{2}{*}{D-Star}   & {\em vBuggy} & 105           & 2              & 2              & 5              & 2,073  (-250\%)     & 2,553   (-195\%)    \\ 
                           & {\em vD4J}  & 105           & 10             & 16             & 23             & 829       & 1,309      \\ \midrule
\multirow{2}{*}{Barinel}   & {\em vBuggy} & 105           & 2              & 2              & 5              & 2,075   (-234\%)    & 2,564  (-185\%)      \\ 
                           & {\em vD4J}  & 105           & 9             & 17             & 28             & 888       & 1,389     \\ \bottomrule
\end{tabular}
}
\end{table}

As a comparison, we also study the effectiveness of fault localization (FL) techniques on {\em vBuggy} and {\em vD4J} for the bugs that belong to Pattern 4 (the fault-triggering tests were not modified). Different from the bugs that belong to Pattern 3, Pattern 4 consists of 150 bugs where the fault-triggering tests were not modified by developers. This implies that the bugs of Pattern 4 may provide a more realistic setting for fault localization. We perform similar analysis on bugs in Pattern 4. 105 out of the 150 unique bugs we analyzed have test failures, while the remaining 45 bugs either have no test failures occurred or the commits before the bug report ({\em vBuggy}) being too old to be retrieved. Surprisingly, we find that the fault localization results for the bugs that belong to Pattern 4 also have become worse. Table~\ref{tab:flPattern4} shows that for Top-1, Top-3, and Top-5, {\em vBuggy} ranges from 2 to 6 while {\em vD4J} ranges from 9 to 28. Similarly, MFR and MAR are much higher for {\em vBuggy} (in the range of 2,000), meaning that developers need to investigate an average of 2,000 to find the faulty statements. The MFR and MAR are in the range of 800 to 1,300 for {\em vD4J}. The decrease in MRF and MAR value is around 200\%.

\rqboxc{{\bf RQ3-Takeaway 1.} The fault localization results on {\em vBuggy} (MFR and MAR around 3,000) are significantly worse than that on {\em vD4J} (MFR and MAR around 700 and 1200). Our findings show that the fault localization results using Defects4J may be significantly different from practical settings.}


\begin{table}
    \caption{The number of bugs with failed fault-triggering tests on the buggy commit ({\em vBuggy}). We consider the bugs whose fault-triggering tests belong to Pattern 3 and Pattern 4 (Table~\ref{tab:patterns}).} \label{tab:nonfailing}
    \resizebox{.6\columnwidth}{!}{
    \begin{tabular}{l|r|r|r|r|r}
    \toprule
        \textbf{Project} & \textbf{Total} & \multicolumn{2}{c}{\textbf{\#Bugs with no}} & \multicolumn{2}{c}{\textbf{\#Bugs with}} \\
        & \textbf{bugs} & \multicolumn{2}{c}{{\bf failed tests}} & \multicolumn{2}{c}{{\bf failed tests}}\\
        && Pattern 3 & Pattern 4 & Pattern 3 & Pattern 4 \\
        \midrule
        Cli & 17 & 0 &  2    &9 & 6  \\
        Closure & 93 & 11 & 3    &32 & 47\\
        Codec & 5 & 1 &  0   &3 & 1  \\
        Collections & 0 & 0 &  0 & 0 & 0   \\
        Compress & 14 & 1 & 2    &8 & 3   \\
        Csv & 2 & 1 & 0      &1 & 0 \\
        Gson & 9 & 1 & 6     &1 & 1   \\
        JacksonCore & 7 & 2 & 3  & 1 & 1   \\
        JacksonXml & 1 & 1 & 0   &0 & 0  \\
        JacksonDatabind & 17 & 0 & 0     &10 & 7   \\
        JxPath & 1 & 1 & 0   &0 & 0  \\
        Lang & 27 & 0 & 0 & 22    &5     \\
        Math & 36 & 2 & 2    &30 & 2   \\
        Mockito & 31 & 1 & 0     &2 & 28  \\
        Time & 3 & 0 & 0     &2 & 1   \\
        \midrule 
        {\bf Total} & 263 & 22  & 18  & 121 & 102
        \\ \bottomrule
    \end{tabular}
    }
\vspace{-0.4cm}
\end{table}

\phead{Discussion.} Since the effectiveness of fault localization techniques degrades significantly, we further investigate the possible causes, other than that the tests contain developer knowledge of the bug. We examine how many bugs whose fault-triggering tests (from Pattern 3 and 4) actually failed in {\em vBuggy} (i.e., the version when the bug was first reported). Note that our study focuses on 263 out of 305 bugs, as some earlier versions of the system could not be retrieved. Table~\ref{tab:nonfailing} shows the number of bugs with failed fault-triggering tests in {\em vBuggy}. 
We noticed that in both Pattern 3 and Pattern 4, a small percentage (15\%) of fault-triggering tests did not cause any failure in {\em vBuggy}, indicating that they were unable to detect the fault.
On the other hand, the remaining fault-triggering tests (85\%) failed and assisted in fault localization.
This result suggests that the majority of these tests did fail in {\em vBuggy}, but they were not effective in detecting the bugs.

Interestingly, in Pattern 4, despite not having fault-triggering tests modified by developers during the bug fixing process, we also observed a significant decrease in the fault localization result for {\em vBuggy}. Through manual analysis of these bugs, we observe that even though developers did not modify the fault-triggering tests, they may still have introduced developers’ knowledge, which can affect fault localization results. Developers may introduce their knowledge in different ways, such as modifying functions involved in test execution or adjusting the test setup configuration. For instance, in the case of bug Lang 57, developers added a {\sf setUp} method to initialize all test execution within the test suite {{\sf LocaleUtilsTest}. The newly added {\sf setUp} method made the fault-triggering test failed if some variables were not correctly initialized before the execution. 
As this modification was made after the bug report was submitted, and no direct modifications was done on the fault-triggering test itself, this change pattern belongs to Pattern 4. However, developers knowledge was introduced in the test setup configuration, which also altered the test execution and coverage, and the results of fault localization. 

Our finding shows that developers’ knowledge may not always be directly incorporated into the fault-triggering tests themselves. Future research may explore bugs from Pattern 4 to investigate and characterize the instances where developers’ knowledge is introduced outside of the fault-triggering tests.
Nevertheless, our findings reveal that fault-triggering tests in Pattern 4, which were not modified by developers during the bug fixing process, resulted in better performance for fault localization, as indicated by both MFR and MAR metrics.
These bugs are better suited for evaluating fault localization techniques in practice. 
Future studies may benefit from using bugs/tests from Pattern 4 to assess fault localization techniques in a more realistic setting.

\rqboxc{{\bf RQ3-Takeaway 2.} Future studies may consider using the fault-triggering tests that were not modified by developers after the bug report to provide a practical setting for evaluating fault localization techniques.}

\section{Discussion and Future Work}
\label{sec:discussion}



In this paper, we studied the bugs and their corresponding fault-triggering tests in Defects4J. We classified the bugs based on whether their tests contain developer knowledge and also the test coverage data on the buggy commit (i.e., the commit prior to the bug report creation). We made the dataset publicly available~\cite{our_repo} and we believe that the dataset can inspire future research. Below, we discuss the implication of our findings and potential future research directions. 


\phead{Future research may use our annotated data to re-evaluate fault localization and automated program repair techniques.}
We find that a majority of the fault-triggering tests in Defects4J contain developer knowledge, and such knowledge can degrade the fault localization results. Our findings provide insights into the effectiveness of fault localization techniques in a more realistic setting where developers may use these techniques in practice (e.g., analyze the failing tests associated with a reported bug). Our dataset also annotates the fault-triggering tests and whether or not they contain developer knowledge. We believe the dataset can be used in three directions. First, future studies can leverage the dataset to re-evaluate the effectiveness of fault localization or automated program repair techniques based on fault-triggering tests that do not have developer knowledge. Second, future research may use our dataset to have a separate evaluation of the bugs (e.g., with and without developer knowledge) in Defects4J. Third, future studies may use our dataset to investigate the characteristics of fault-triggering tests that require less maintenance during bug fixes, such as tests that involve little to no code changes (e.g., tests in Pattern 4).

\phead{Future studies may need to consider developers' bug-fixing process when creating benchmarks.}
In RQ2, we found that fault-triggering tests may not be available when developers receive a bug report. Developers may then develop fault-triggering tests while working on a bug fix to replicate the bug. 
Hence, our findings call for more empirical studies to further understand developers' bug-fixing activities, and how to consider such activities when creating benchmark datasets. 
Future studies may use our dataset as an initial step to understand the bug-fixing process and expand the study on other systems and settings. It is important to note that our findings particularly resonate in scenarios where developers heavily rely on existing test suites to diagnose and fix bugs. For software projects where automated testing is a primary method for quality assurance and where developers often modify or add tests post-bug-fixing, our insights can guide the process of benchmark dataset creation and assessment.




\phead{There is a need for more comprehensive and realistic benchmarks.}
To help facilitate software testing research, researchers have proposed several bug benchmarks. Among all, Defects4J~\cite{just2014defects4j} is the most popular and widely used benchmark. 
Although Defects4J provides a clean dataset that is easy to use, some of the bugs and tests are manually crafted or contain data from ``future'' commits. As demonstrated in this paper, some fault-triggering tests may contain developer knowledge of the bug, leading to an overestimation of fault localization results.
Apart from Defects4J, other datasets like Bugs.jar~\cite{saha2018bugs} and BEARS~\cite{madeiral2019bears} may also be impacted by our findings. For example, Bugs.jar follows a similar approach to Defects4J, where bugs and tests are manually isolated based on "future" commits. Through our initial manual examination, in BEARS, some of the fault-triggering tests were also collected from ``future'' commits. These datasets share similarities with Defects4J and may exhibit similar concerns related to developer knowledge in fault-triggering tests. We acknowledge the challenges with curated datasets and their limited generalizability. However, it is crucial to understand the specific biases and assumptions in these datasets. While our findings on datasets like Defects4J, Bugs.jar, and BEARS may not be surprising to seasoned researchers, our detailed analysis serves as a reminder and a guide for newer researchers or practitioners in the field. We urge the community to consider the assumptions and constraints under which these datasets were created and to exercise caution when generalizing their findings.
As a result, we believe there is a need for a more comprehensive and realistic benchmark. As an example, a recent study~\cite{chen2022t} compiles and executes a consecutive sequence of commits in multiple systems and collects the test coverage. In a way, such data provides a more realistic setting for evaluating fault localization and automated program repair techniques. In short, we believe there is a need for more comprehensive and realistic benchmarks for software testing research.


\section{Related Work}\label{sec:related}
\phead{Empirical studies on Defects4J.}
With the popularity of automated debugging solutions, the Defects4J benchmark has been widely used in evaluating the effectiveness of research approaches.
However, only a few studies~\cite{koyuncu2019ifixr, liu2021critical} have pointed out the presence of future tests.
~\citet{koyuncu2019ifixr} proposed a bug report-driven program repair technique in their study and noted that tests that were modified during bug fixes may impact the evaluation of their technique.
Some of the authors of this work made the same observation in their following work~\cite{liu2021critical} when evaluating the biases in automated program repair benchmarks. However, there has been no systematic study on the prevalence of future tests, and prior studies did not evaluate the impacts of these tests on fault localization techniques.
In our study, we conducted a comprehensive analysis of the timeline of fault-triggering tests and their evolution over time. Our findings indicate that fault-triggering tests may undergo changes throughout the entire bug resolution process, starting from bug creation to its resolution. We also studied the impact of these modified tests on fault localization. 
~\citet{just2018comparing} also discussed some limitations of fault-triggering tests in Defects4J. ~\citet{just2018comparing} found that some tests may not accurately reflect the information developers have in real-world scenarios. 
However, the prevalence of developer knowledge in these tests has not been thoroughly investigated. 
In our paper, we presented a manual study on how developers modified these tests, and evaluated their impact on the performance of state-of-the-art spectrum-based fault localization techniques.

\phead{Bug benchmarks.}
Several bug benchmarks have been proposed by the automated testing research community to support empirical evaluations. Besides Defects4J, two other popular Java benchmarks are Bugs.jar~\cite{saha2018bugs} and BEARS~\cite{madeiral2019bears}.
Both of these benchmarks follow the same bug replication process as Defects4J, where fault-triggering tests may be committed into the versions of the system before bug fixes.
However, through our initial investigation, some of these tests were added after the bug report or during the bug fix, which means they may also suffer from a similar issue as Defects4J. Future studies are needed to study the timeline of the fault-triggering tests in other benchmarks and what is the impact on the downstream research approaches. 



\phead{Fault Localization and Program Repair Techniques.}
Due to the importance and high cost of bug-fixing activities, a large number of research studies focus on fault localization (FL)~\cite{b2016learning, TraPT, FLUCCS, MULTRIC, zheng2016fault, DeepFL} and automated program repair (APR)~\cite{icse20,li2022dear,saha2017elixir}. The Defects4J dataset has been used extensively as a benchmark in the evaluation of FL and APR techniques.



\noindent\textbf{\em Fault Localization (FL).} Since its initial release in 2014, the Defects4J dataset has been widely used for evaluating fault localization techniques.
Learning-to-Rank (LtR) technique has been used to improve fault
localization~\cite{b2016learning, TraPT, FLUCCS, MULTRIC}. The basic idea is to learn the potential faulty locations via combining suspiciousness values computed by various fault localization techniques (i.e., features), then adopt LtR to rank the suspicious elements. MULTRIC \cite{MULTRIC} combines different suspiciousness values from spectrum-based fault location (SBFL). Some work combines SBFL suspiciousness values with other information, e.g., program invariant \cite{b2016learning} and source code complexity information \cite{FLUCCS}, for more effective LtR in FL.
Orthogonal to coverage utilization, deep learning-based approaches~\cite{lou2021boosting, DeepFL, cnn-fl, zhang2019cnn, icse21-fl} learn embeddings of test coverage information to generate the suspicious score of code elements. For example, GRACE~\cite{lou2021boosting} proposes a novel graph representation that combines method and test coverage information, and learns to rank the faulty methods. 

All of the above mentioned techniques run the test cases and source code of Defects4J to collect the test coverage information and utilize different techniques to calculate the suspiciousness scores for the relevant code elements. 
In this work, our objective is to evaluate the influence of developers' knowledge on the performance of state-of-the-art SBFL techniques. This assessment can serve as a baseline for other coverage-based fault localization techniques.

\noindent\textbf{\em Automated Program repair (APR).} 
Defects4J is also extensively and widely used in the evaluation of APR techniques~\cite{wen2018context,liu2019you,jiang2019manual,tbar-issta19}. 
One common approach in the literature to evaluate the correctness of the patches is to verify that the candidate patch makes the original program pass all test cases. 
Our work investigates the prevalence of developers' knowledge in tests, which may impact prior APR research that was performed on Defects4J.
In addition, fault localization is a crucial step in the APR research. It helps to reduce the search space for generating patches. The test cases, particularly the fault-triggering tests that cover the buggy locations, play a crucial role in determining which part of the code should be repaired.
Some prior studies~\cite{yang2017better,xiong2017precise} also use existing test cases to eliminate overfitted patches and improve the quality of generated patches, which may also be impacted by future tests.
\section{Threats to Validity}
\label{sec:threats}

\noindent{\bf Internal Validity.} Threats to internal validity are related to experiment errors or biases. One main threat comes from the human analysis in our study. We manually analyzed and categorized the modifications on fault-triggering tests in RQ2. 
However, we adopted the manual analysis approaches used in prior studies~\cite{li2019dlfinder, lamothe2020apis, ding2020towards, li2020shall} to mitigate subjectivity. Three phrases were used and two authors independently were involved in the analysis. The analysis results achieve a Cohen’s Kappa of 0.86, which indicates a substantial level of agreement~\cite{cohen1960coefficient}. 
In RQ1, we also manually reviewed the relevance of the test modification to a bug. However, the main work in extracting events in relation to bug resolution was automatic and we only manually check whether any errors in the generated results.


\noindent{\bf External Validity.} Threats to external validity concern the degree to which our findings can be generalized. While our study primarily centers on the Defects4J dataset, it is crucial to recognize that other datasets with test cases and bug information might yield different outcomes. The Defects4J dataset, being a predominant benchmark for significant software engineering tasks like fault localization and automated program repair, has deeply ingrained its relevance in the research community. Numerous studies, particularly in the APR domain, lean heavily on the Defects4J dataset for evaluating Java programs. Given its widespread use, it is conceivable that some characteristics of Defects4J might have influenced research results or methodologies. However, certain key insights from our analysis, such as the impact of developer knowledge in fault-triggering tests and the observed modification categories, are not inherently tied to Defects4J. Thus, they could potentially manifest in other datasets as well. For example, datasets derived from projects with similar development practices or that undergo similar processes of curation could exhibit comparable biases or characteristics. Furthermore, it's worth mentioning that while the core of our findings is rooted in Defects4J, the underlying themes—like the influence of developer knowledge—could be pervasive in other datasets. As such, researchers exploring datasets akin to Defects4J should consider the potential implications and biases our study unveils.


\section{Conclusion}\label{sec:conclusion}

A large amount of research effort focuses on assisting in locating faults in the source code.
Fault localization and automated program repair have been widely studied to provide developers with an automated debugging solution.
Various benchmarks, like the popular Defects4J, assist in evaluating fault localization techniques. It contains real software bugs, relevant fault-triggering, and bug fixes. However, the fault-triggering tests in Defects4J might be extracted from a version of the system where the bug is already fixed. Thus, these tests may offer clues to fault localization techniques regarding the bug's location.
We present a case study that examines developer knowledge in fault-triggering tests. Our results show that 77\% of the fault-triggering tests in Defects4J contain developer knowledge. Our manual study highlights that developers often add new tests/assertions to replicate the bug or modify existing tests for regression testing, leading to the existence of such knowledge.
We also conduct an experiment on the impact of tests with developer knowledge on the performance of state-of-the-art spectrum-based fault localization techniques. Our experimental results show that when evaluating tests without developer knowledge, the performance of fault localization techniques significantly degrades. This is because the fault-triggering tests, which no longer contain developer knowledge, do not fail and do not contribute to locating the fault.
Finally, we provide some potential research directions that may be done using our dataset.
In our study, we highlight important findings on the developer knowledge that may exist in fault-triggering tests, overestimating the performance of fault localization research in a realistic setting. We present our dataset publicly available that may be used for new or extension to existing benchmarks. 

\bibliographystyle{ACM-Reference-Format}
\bibliography{reference, FLref, APRref}

\end{document}